\documentclass[journal]{IEEEtran}
\usepackage[usenames]{color}
\usepackage{epsfig}
\usepackage{graphics}
\usepackage{amsmath}
\usepackage{amssymb}
\usepackage{multirow}
\usepackage{cite}
\usepackage{array}
\usepackage{pslatex} 
\usepackage{url}
\usepackage{caption}
\usepackage{lineno}
\usepackage{graphicx}  
\usepackage{setspace}
\usepackage{tikz}
\usepackage{hhline}
\usepackage{mathtools}
\usepackage[letterpaper]{geometry}
\ifCLASSINFOpdf
\else
\fi
\usepackage[english]{babel}
\usepackage[utf8]{inputenc}
\usepackage{fancyhdr}
\begin{document}
%
\title{\textcolor{violet}{{AI-Driven Analysis of the Effects of Recreational Activities to Well-Being using Physiological Responses: A survey}}}

%
%
%

\author{Aditi Site,
        Tarmo Lipping,~\IEEEmembership{Senior Member,~IEEE}
\thanks{First Author is with Data Science Research Centre, Tampere University, Tampere, Finland (e-mail: aditi.site@tuni.fi). }
\thanks{Second Author is with Data Science Research Centre, Tampere University, Tampere, Finland (e-mail: tarmo.lipping@tuni.fi).}}

\maketitle\thispagestyle{fancy}

\begin{abstract}
Well-being is a broader concept ensuring psychological, social, physical, and cognitive health. Recreational activities are one of the major approaches to facilitate well-being. There are various recreational activities in each of the domains in which well-being can be measured. Although well-being is often assessed using subjective scales, physiological responses such as heart rate, heart rate variability, skin responses and brain activity data provide objective ways to estimate the level of well-being. Research has focused on using these measurable physiological signals to develop AI driven well-being analysis concepts. This study presents a review of previous studies that have considered analyzing well-being. This review summarizes studies on the basis of recreational activities, data collection scenarios, subjective and objective data, data analysis methods and evaluation criteria. Furthermore, this study also proposes a framework to enhance data collection, analysis and visualization setup for the assessment of well-being.
\end{abstract}


%
\IEEEpeerreviewmaketitle

\textbf{\textit{Take-Home Message---} The development of AI driven technologies would enhance user engagement and contribute to well-being by facilitating mental health interventions. This will also help in designing of the recreational activities catering to specific needs (emotional well-being, stress reduction etc.)}\\
\\

\newpage
\section{INTRODUCTION}
\IEEEPARstart{W}{ell}-being refers to the positive state of living and maintaining an overall quality of life in terms of  social, cognitive, physical and mental health. Well-being depends on avoiding negative stimuli and feeling good psychologically, emotionally and spiritually \cite{wb_ra1}. With growing mental health issues leading to chronic illness, achieving overall well-being has become a central concern. Overall well-being can be achieved by distancing from the negative stimuli through various recreational activities. Recreational activity refers to leisure pursuits engaged in for enjoyment and relaxation as well as fun and fitness. With the aging population, the risk of physical health issues such as chronic illness and mental health issues such as loneliness, boredom, depression, cognitive decline or anxiety increases. This can have worsening effect on the overall well-being of the person \cite{wb_ra2}. Recreational activities have benefits in improving the health and overall well-being of the person. 

Technology plays a crucial role in delivering solutions for achieving and assessing overall physical and mental well-being. Wearable sensors have the potential to monitor continuously the crucial vitals for the physical and mental health. Tracking physiological parameters such as heart rate, heart rate variability, breath rate or skin response can be utilized to analyze the presence of stress and anxiety. Similarly, wearable sensors can also be used to track the behavioral parameters such as sleep quality, physical activity, social interactions or mobility patterns that can help in estimating the level of mental health \cite{wb_ra3}. Some of the physiological variables that can be used to track physical and mental health include Electroencephalogram (EEG), Electrocardiogram (ECG), Galvanic skin response (GSR) or Electro-dermal activity (EDA), Photoplethysmogram (PPG), Respiration (Resp) or Mobility (assessed using Accelerometers (ACC) or Gyroscopes (Gyro)). Wearable sensors used to acquire these variables provide a non-invasive way to monitor and record the crucial body vitals such as heart rate, heart rate variability, skin conductance, brain activity, movement patterns etc. 

Analytical methods using machine learning analysis have been widely used recently to analyze the data from these sensors. Traditional machine learning models such as support vector machines (SVM), bagging and boosting methods along with sophisticated deep learning models have been used to derive and analyze the sensor data and provide significant outcomes related to physical and mental health \cite{wb_ra4}. This has led to developing techniques using Artificial Intelligence (AI) to facilitate overall well-being.

In this study, our objective is to explore the potential of AI-driven methods to facilitate well-being using physiological data. We aim to explore the following research questions:
\begin{enumerate}
    \item How to define/measure well-being?
    \item What recreational activities can help improve well-being?
    \item What factors affecting well-being can be measured using sensor data?
\end{enumerate} 
\begin{figure*}[h]
  \includegraphics[width=\textwidth]{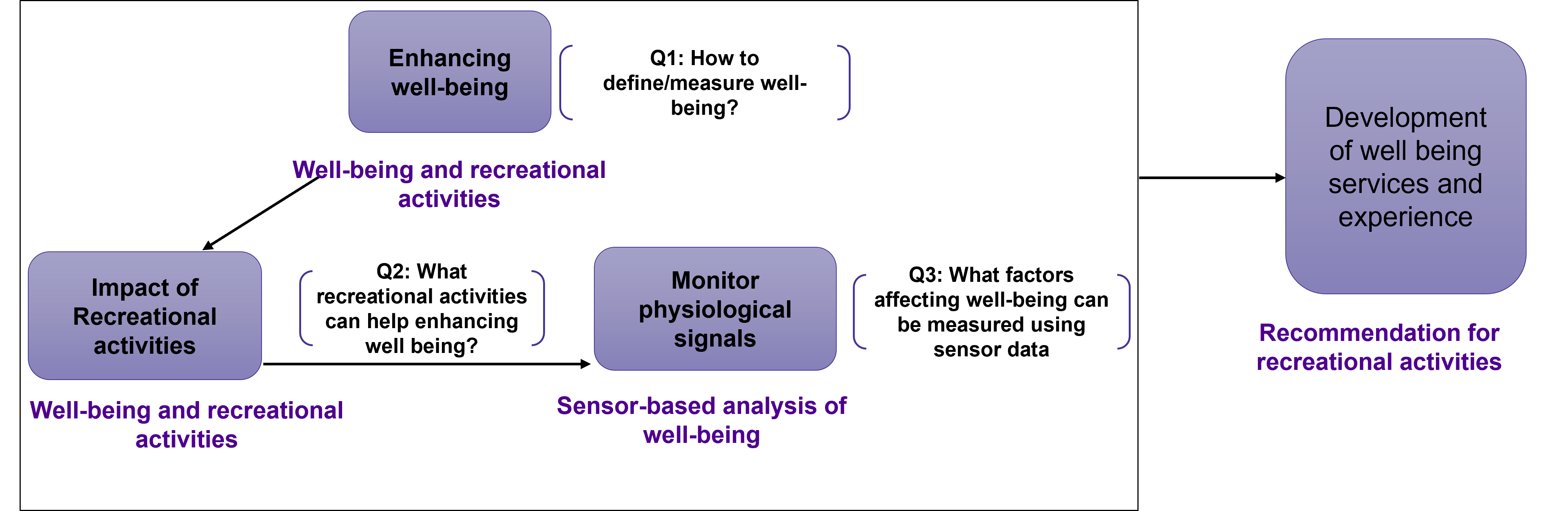}
  \caption{Research question map}
  \label{Fig:researchquestionmap}
\end{figure*}
Figure \ref{Fig:researchquestionmap} illustrates the research questions with the logical association between the conceptual keywords such as well-being, recreational activities and physiological responses. The first research question is associated with the concept of well-being and its purpose is to identify the components that define and measure various aspects of well-being. The second research question considers the recreational activities that improve the well-being of a person. This research question also focuses on identifying component-wise recreational activities for well-being. The third research question deals with physiological signals and variables acquired using wearable sensors that can be used for modeling the indicators of well-being. This question also explores the machine learning and deep learning techniques used for affect modeling. Monitoring physiological signals during recreational activities facilitates behavioral analysis and enables modeling the mental states indicating a person's well-being. However, such modeling could also be used to recommend certain recreational activities based on behavioral analysis of a person. This study also considers the possibility of developing a recommendation system that can enhance the well-being effect of recreational services.

\section{Conceptual background}
The concept of well-being is evolving and the development of sensor technologies and machine learning algorithms has led to a new dimension involving AI-driven methods for analyzing and identifying person's well-being. Additionally, recreational activities play a key role in analyzing well-being of a person. Many studies have focused on analyzing the physiological responses collected during recreational activities to assess the overall well-being of a person. 

\begin{figure*}[h] 
    \includegraphics[width=\textwidth]{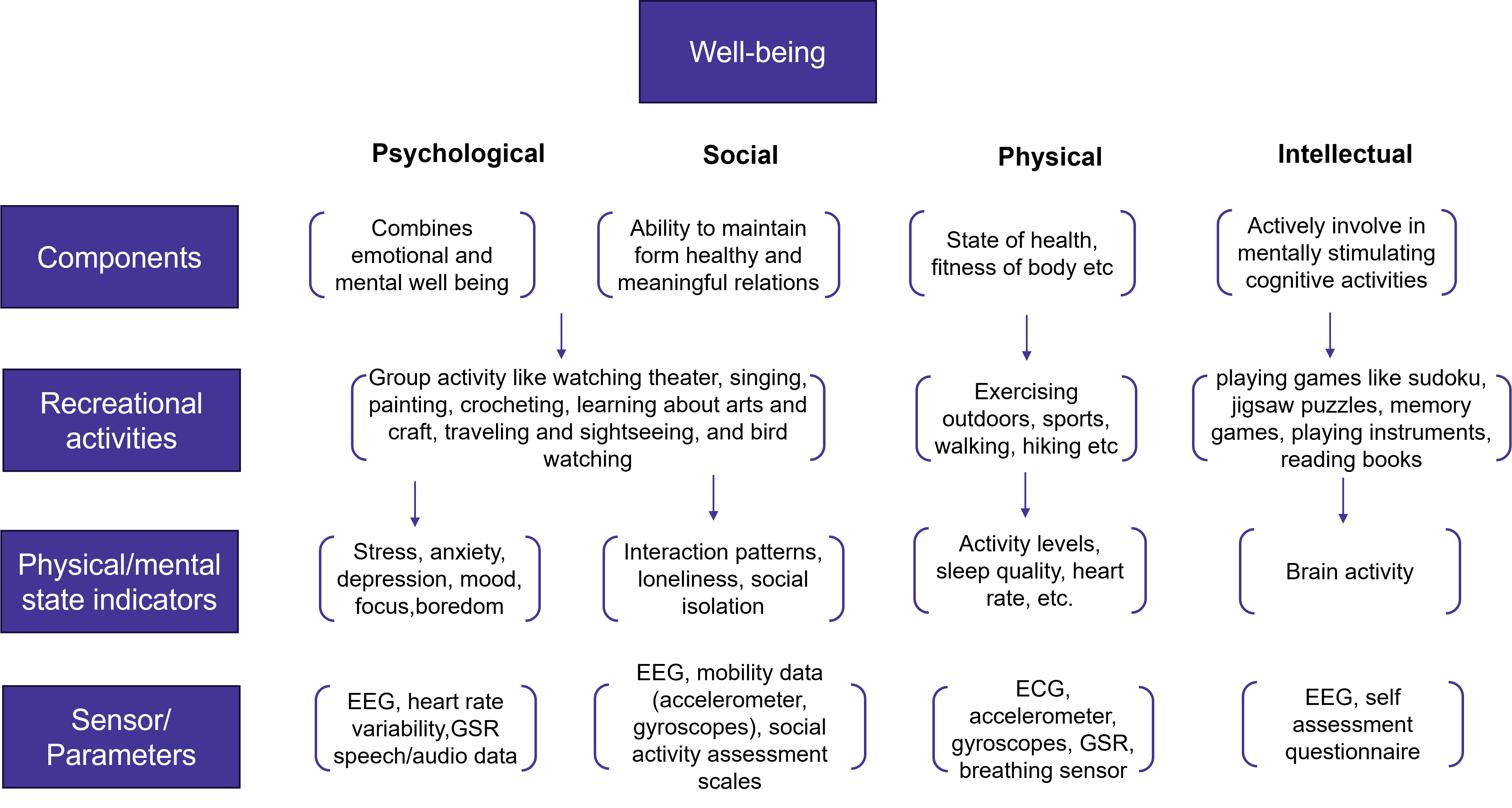}
    \caption{Key concepts and their relationships}
    \label{fig:conceptrelationship}
\end{figure*}

This section provides an overview of the key concepts related to well-being, recreational activities, and physiological responses, as well as the relationships between them. Figure \ref{fig:conceptrelationship} highlights the three key concepts and how they are related. Detailed explanation of the concepts with respect to Figure \ref{fig:conceptrelationship} is given in the following subsections.

\subsection{Well-being}
Well-being is a multidimensional concept and improving overall quality of life depends on promoting cognitive, physical and mental well-being \cite{cb_well-being}. Figure \ref{fig:conceptrelationship} shows various components that define overall well-being. These components are similar to those presented in the perceived wellness model of \cite{wb_ra1}. In the perceived wellness model, there are six components, namely psychological, social, intellectual, emotional, spiritual, and physical. In the scope of our study we have considered only those components that can be measured using physiological responses such as psychological, social, physical, and intellectual (see Figure \ref{fig:conceptrelationship}). Psychological well-being refers to the emotional and mental state of a person and the presence of positive psychological functioning. It refers to an individual's ability to cope with challenges and maintain a positive attitude. Social well-being refers to the individual's ability to maintain meaningful relationships and integration with the community. Physical well-being refers to the overall health and the functioning of the body. It includes the ability to perform daily tasks without physical limitation and maintaining overall fitness of the body. Intellectual well-being or cognitive well-being refers to a person's ability to get involved in mentally stimulating tasks and think creatively and critically. This emphasizes adapting to challenges and solving complex problems. All these components comprehensively evaluate overall well-being of a person. However, there are certain recreational activities corresponding to each component that can help promoting well-being. These activities are discussed in the next subsection.

\subsection{Recreational activities}
Recreational activities are the leisure activities that a person voluntarily takes to engage in enjoyment and relaxation and to take breaks from daily routines. These activities help to promote well-being in various dimensions. Figure \ref{fig:conceptrelationship} shows various recreational activities in psychological, social, physical, and intellectual dimensions. These activities, engaged in either individually or in groups, include art and culture activities, watching and/or performing in theater groups, singing, painting, traveling, nature walks etc. They help in social bonding and stress reduction, hence promoting psychological and social well-being \cite{cb_well-being}. Certain outdoor activities such as exercising outdoors or engaging in sports help promoting physical well-being. Studies in \cite{wb_ra1} suggest that extreme sports contributes to recreational flow experiences. Similarly, cognitive recreation can be stimulated through activities involving mind games such as puzzles, sudoku, memory games, reading books, playing instruments etc. This helps promoting intellectual or cognitive well-being. 

\subsection{Physiological responses}
Being a multi-dimensional concept, measuring well-being involves assessing it objectively and subjectively. Subjective assessment involves various self-assessment scales such as the Perceived Wellness Scale \cite{wb_ra1}, Positive and Negative Affect Scale (PANAS), State-Trait Anxiety Inventory (STAI) for anxiety level or Self-Assessment Manikins (SAM) \cite{wesad} while objective measurement involves data acquired using various sensors corresponding to the various components of well-being. These measurable data come from sensors such as EEG, ECG, PPG, EDA/GSR or Accelerometers that can measure bodily responses. From the sensor data, physiological variables such as heart rate, heart rate variability, skin conductance, movement data, brain activity data etc. can be derived. Figure \ref{fig:conceptrelationship} shows various physical and mental state indicators and the sensors that can be used to predict/assess these states. 

\begin{figure}
  \includegraphics[width=\columnwidth]{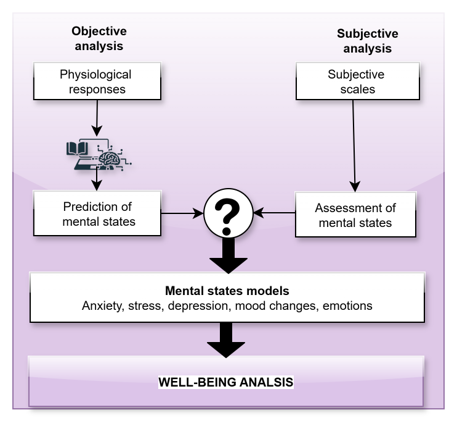}
  \caption{Relationship between data and well-being}
  \label{fig:relationship_data_wellbeing}
\end{figure}

The relationship between the different types of data and mental states corresponding to well-being is illustrated in Figure \ref{fig:relationship_data_wellbeing}. While physical well-being is more straightforward to measure objectively, the other components of well-being (psychological, social and intellectual) are usually assessed by predicting mental states such as anxiety, stress, depression, mood or emotions. To estimate these mental states based on measurable physiological responses, machine learning models are commonly used \cite{detecting_anxiety,HRV_biomarker,detecting_stress}. Developing and training the models is based on subjective assessment of these mental states using questionnaire-based scales. 

While presenting a general concept, there are several open issues related to the framework of Figure \ref{fig:relationship_data_wellbeing}. These issues are related to the relationships between the blocks of the schematic such as:
\begin{itemize}
    \item{How well do the various subjective scales actually describe the mental states?}
    \item{What is the relation between mental states and the various aspects of well-being?}
    \item{How to deal with the inter-individual variability in the subjective experience of well-being?}
\end{itemize} 
This study aims at clarifying the concepts and providing tools to explore these issues more systematically in the future. 

As an integrated whole, well-being analysis necessitates designing of a setup that involves recreational activities, wearable sensors for measuring physiological responses, data analysis and modeling framework as well as subjective scales for training and validating these data-driven models. 

\section{Methodology and organization}
This section outlines the methodology used to investigate state-of-the-art methods in the field of the assessment of well-being using physiological responses. A narrative review is presented on performing AI-driven analysis of physiological responses such as EEG, ECG, GSR, PPG etc. to explore the effect of recreational activities on a person's well-being. A comprehensive literature search was conducted primarily using the IEEE Explore database. Keywords including "well-being", "recreational activities", "physiological responses", "machine learning", and "behavior analysis"  were used in various combinations. Only English-language peer-reviewed articles were considered. Articles were selected based on their relevance to the key themes of this review.

\begin{figure}
  \includegraphics[width=0.5\textwidth]{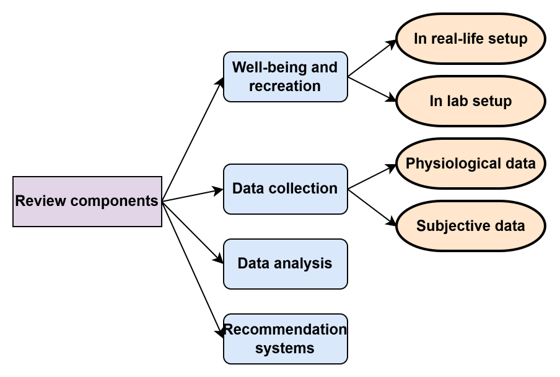}
  \caption{Review organization}
  \label{fig:review_organization}
\end{figure}

Figure \ref{fig:review_organization} shows the structure of the review. This review is divided into 4 components and further divided into specific sub-components. The first component, well-being and recreation, focuses on two types of data collection scenarios that can be used to study individual's well-being. Data collection in real-life situations and in the lab environment is considered. The second component, data collection, which is further subdivided to consider physiological data and subjective data, explains the data collection methodology and the type of data collected. The third component explores the different types of qualitative, quantitative and machine learning analysis techniques being applied to the physiological and subjective data. This component also focuses on machine learning models as well as various algorithms and evaluation criteria used for analyzing the physiological responses. The last component, recommendation systems, details on the possibility to develop a comprehensive tool for recommending recreational activities and evaluating their effect on well-being. The comprehensive analysis of literature under the specified components is presented in section \ref{section:reviewdetails}.

\section{Comprehensive analysis of methods} \label{section:reviewdetails}
This section provides comprehensive review of studies related to the four main components shown in Figure \ref{fig:review_organization}. The objective of the review is to examine the existing literature and, based on the outcomes and gaps, develop a methodology to model well-being using AI-driven analysis. This section is divided into four sub-sections, each analyzing one of the key points in the assessment of well-being.

\subsection{Well-being and recreation}
Two different setups are considered, real-life environment and the lab setup. In real-life environment, the well-being analysis is usually done when the person is involved in some recreational activity to trigger their emotions. These kind of situations include theater performances, concerts or visiting art exhibition, for example. In the lab setup the subject is usually sitting in a quiet room and is presented with stimuli such as movie excerpts, audio or images. Alternatively, the subject may be engaging in a well-controlled activity such as a cognitive, mental or physical test. In the lab, the stimuli are usually better controlled and the effect of external sensory input is minimized.

\subsubsection{\textbf{In real-life environments}}
\begin{table*}
\caption{Well-being and recreation (In real-life environments)} 
\label{tab:wellbeing and recreation real-life}
\footnotesize\centering
\begin{tabular*}{\textwidth}{@{\extracolsep{\fill}}|p{0.2in}|p{0.6in} |p{0.8in}|p{1in}|p{1in}|p{1in}|p{1.3in}|}
  \hline
  \textbf{Ref.}  & \textbf{Subjects} & \textbf{Environment} & \textbf{Data collected} & \textbf{Sensors used} & \textbf{Analysis method} & \textbf{Results and evaluation} \\
  \hline
  \cite{wb_ra1}  & Sports participant& Extreme sports rock climbing, paragliding, scuba diving & Questionnaire based scales& -  & Quantitative (SPSS), regression analysis & Correlation scales showed negative and positive correlation with wellness indicators (\textbf{Target}: Encounter, boredom, anxiety, flow) \\
  \hline
  \cite{cb_well-being} & Older adults& Retirement community & Motion and audio data along with questionnaires based scales & Accelerometer, barometer, light, temperature, humidity sensors, compass and microphone & Machine learning analysis using HMM (Hidden markov Model) and ensemble algorithms) & Accuracy: 83.7\%, Precision: 90\%, Recall: 84\%  (\textbf{Target}: Social isolation, mood, depression, physical activity index\\
  \hline
  \cite{wb_ra5}  & Adults (20-45 years) & Outdoor activities &Experience sampling questionnaires & GPS (Global positioning service), Accelerometer & Quantitative analysis & Feasilbility analysis for smartphone based technologies to capture activity patterns and perceived emotions (\textbf{Target}: Human activity patterns, emotions)\\
  \hline
  \cite{wb_ra6} & Older adults with dementia& Robot assisted recreational activity & Speech data & Voice recorder&Graphical and descriptive analysis with basic visualization &Duration of speech analyzed (\textbf{Target}: Happy, joy, excitement) \\
  \hline
  \cite{wb_ra7} & Adolescents & Daily living (Studying, smartphone using, metaverse gaming)& Smartphone data & accelerometer, gyroscope, ambient light, usage data & Customizable Automated Machine Learning Process used &Accuracy range 74\%--86\% for detecting various disorders (\textbf{Target}: Gaming disorder, anxiety, depression, impulsiveness, aggression, Attention deficit hyperactivity disorder (ADHD), Obsessive-Compulsive disorder (OCD), Prodromal) \\
  \hline
\end{tabular*}
\end{table*}

Table \ref{tab:wellbeing and recreation real-life} summarizes the studies focusing on identifying person's well-being in different real-life situations. Information is organized into columns according to the participating subjects, the recreational activity involved, the environment where the data were collected, the type of data collected, the sensors used, the analysis method applied and the evaluations or outcomes of the study. It can be observed that the previous studies have been performed on a diverse group of participants/subjects including younger and older adults along with specific groups such as sports participants. Also, the recreational activities during which the data were collected reflects psychological, social (retirement community), physical (extreme sports) and cognitive (robot assisted recreation) components. The data collected from the participants includes questionnaires, i.e., subjective data. In \cite{wb_ra1}, researchers have used Personal Information Form (PIF), Alpak Flow Scale (AFS) and Perceived Wellness Scale (PWS) to record participants' socio-demographic characteristics, flow experiences during recreational activity and wellness level, respectively. Researchers in \cite{wb_ra5} have used Experience Sampling Methods (ESM) which consists of self-reports of experiences by the participants. Studies \cite{cb_well-being, wb_ra6} have also recorded motion and speech data to analyze the interaction of participants with each other and robots, respectively. Furthermore, multiple wellness scales such as Friendship Scale to measure social isolation and connectedness, SF-36 to evaluate overall well-being, Center for Epidemiological Study Depression Scale (CES-D) for depression screening and Yale Physical Activity Survey (YPAS) to measure physical activity parameters have been used in \cite{cb_well-being}. When data analysis is considered, most of the studies have used either graphical representation or qualitative/regression analysis except in \cite{cb_well-being} which has used machine learning analysis. This could be because in most of the studies hypothesis testing is done to evaluate the possibility of using different methods such as smartphones \cite{wb_ra5} or robots \cite{wb_ra6} to examine the well-being. Likewise, in \cite{wb_ra1}, the relationship between the recreational flow experiences and the perception of well-being is analyzed during extreme sports. 

\subsubsection{\textbf{In lab environments}}
\begin{table*}
\caption{Well-being and recreation (In lab environments)} 
\label{tab:wellbeing and recreation lab}
\footnotesize\centering
\begin{tabular*}{\textwidth}{@{\extracolsep{\fill}}|p{0.2in}|p{0.6in} |p{0.8in}|p{1in}|p{1in}|p{1in}|p{1.3in}|}
  \hline
  \textbf{Ref.}  & \textbf{Subjects} & \textbf{Environment} & \textbf{Data collected} & \textbf{Sensors used} & \textbf{Analysis method} & \textbf{Results and evaluation} \\
  \hline
  \cite{wb_ra_l4} & Students & K-emocon dataset used (Lab setup: 16 sessions of 10 min debate between subjects) & Empatica E4,Polar H7 heart rate sensor,Neurosky mindwave headset & Accelerometer, Photoplethysmograph(PPG), EDA, Heart rate, ECG, EEG, Temperature&Machine learning analysis (Decision tree, Random Forest, Multi-Linear, polynomial, XG boost, Ridge, and lasso model)& Random Forest regressor performs best with an R2 score of 0.84 and MSE of 0.27 (\textbf{Target}: Emotions)\\
  \hline
  \cite{wb_ra_l5} & Adults & CASE dataset used (Lab setup: 8 videos for 4 emotions were used) & Thought technology sensor& EMG (Electromyogram), Blood volume pulse (BVP),ECG, Temperature, Resp& Machine learning analysis (clustering (K means), distance calculation, Linear regression, Neural network and fuzzi logic) & The measures of errors are given by MSE and RMSE (\textbf{Target}: Emotions such as Good Mood, Sadness, Anger, Stress, and Impatience)\\
  \hline
  \cite{wb_ra_l1} & Young adults (average age 19 years)& Lab setup (Impromptu speech task)& E4 Empatica watch & EDA, Skin temperature (ST), heart rate sensors&Machine learning analysis (SVM, Decison Tree, Random forest, K-Nearest Neighbour)& Accuracy between 97.54\% and 99.48\% (\textbf{Target}: Social anxiety)\\
  \hline
  \cite{wb_ra_l2} & Children with Autism& Lab setup (Walking on treadmill)& Shimmer 2r sensors & Heart rate and accelerometer data&Multiple model Kalman filter& Accuracy  93\% (\textbf{Target}: Arousal in autistic children)\\
  \hline
  \cite{wb_ra_l3} & Graduate students & WESAD dataset used (Lab setup: Used video clip, Stress test, meditation, rest)& RespiBAN, Empatica E4 & ECG, EMG, EDA, Temperature, Accelerometer &Machine learning analysis (Boosting algorithms)& Accuracy  95\% and 99\% (\textbf{Target}: Stress)\\
  \hline
  \cite{wb_ra_l_survey2} & Survey on different age groups & Daily life environment including office, car, home, campus& Smartphone and wearables & Heart rate, EDA, PPG, EEG & Machine learning analysis & Accuracy between 70\% and 95\% (\textbf{Target}: Stress)\\
  \hline
\end{tabular*}
\end{table*}

Table \ref{tab:wellbeing and recreation lab} summarizes the studies that have been performed in lab settings to assess various emotions and conditions such as anxiety or stress, and analyze the physiological responses to predict mental states such as emotions, stress, anxiety, depression etc. Some of the studies \cite{wb_ra_l4, wb_ra_l5, wb_ra_l3} have used publicly available datasets for emotion recognition involving watching video clips related to various emotions (detailed information on such datasets is presented in Table \ref{tab:datasetreview}). However, there are also studies that have used stress tests, cognitive tasks or physical tasks to record the data corresponding to the wellness indicators. In \cite{detecting_anxiety}, impromptu speech tasks were used to analyze the social anxiety among young adults. Similarly, in \cite{wb_ra_l2} researchers have analyzed the anxiety among the austistic children using simple walking tasks. Further surveys published on stress detection have used different stress-inducing environments (such as office environment or automobile environment) or stress-inducing tests (such as social stress test, cold pressor test, Montreal imaging tests etc.) and studied how different physiological responses (such as heart rate, body temperature, skin response) behave due to stress \cite{wb_ra_l_survey2}. The common physiological responses recorded in all these studies were mobility, PPG, EDA, heart rate, temperature and EEG, and the common devices used to collect these data were smartwatches such as Empatica E4 or sensors such as Shimmer, RespiBAN, EEG headsets etc. In addition to objective sensor data, these studies have also used some subjective scales to establish the ground truth. Data analysis is most commonly performed using machine learning methods.

From the above two subsections and Tables \ref{tab:wellbeing and recreation real-life} and \ref{tab:wellbeing and recreation lab}, it can be observed that previous studies have used different methodologies to measure the well-being in real-life and lab environments. For real-life environments mostly subjective methods, including questionnaire-based scales, are used whereas in lab environments objective data measured using smartwatches or other wearables are more common. Furthermore, the analysis methods also differ for different types data. Subjective data are commonly analyzed using quantitative methods, whereas machine learning methods are applied to objective data. However, common physiological responses including movement, temperature, PPG, EEG, heart rate and EDA are used in both type of situations and there exists significant potential in modeling the well-being indicators using these responses. In addition, it is also possible to develop emotion recognition models using these physiological responses in real-life environment.

\subsection{Data collection}
\begin{table*}
\caption{Review of available datasets} 
\label{tab:datasetreview}
\footnotesize\centering
\begin{tabular*}{\textwidth}{@{\extracolsep{\fill}}|p{0.2in}|p{0.6in} |p{0.8in}|p{1in}|p{1in}|p{1in}|p{1.3in}|}
  \hline
  \textbf{Ref.}  & \textbf{Subjects information} & \textbf{Data Collection Scenario} & \textbf{Device used} & \textbf{Sensors used} & \textbf{Annotations/Emotions} & \textbf{Subjective scales} \\
  \hline
  \cite{wesad} & 15 subjects (Graduate students), Average age: 27.4 years & Lab settings using video clip, stress test, meditation, rest. Duration: 2 hours & RespiBAN, Empatica E4 & ECG, EDA, EMG, TEMP, BVP,  Acc& Neutral, stress, amusement& PANAS, STAI, SAM, SSSQ\\
  \hline
  \cite{dreamer} & 23 subjects, Age range: 22-33 years & 18 video clips (clips duration 1-6 min, Avg: 3.3 min). Duration: 1.5-2hours & Emotiv EPOC, Shimmer ECG & ECG, EEG& Valence, arousal, dominance & SAM\\
  \hline
  \cite{k_emocon} & 32 subjects (Students), Age range: 19-36 years & 16 sessions of 10 min debate between subjects. Duration: 3hours & Empatica E4, Polar H7 heart rate sensor, LooknTell head mountcamera, Neurosky mindwave headset& ACC, PPG, EDA, Heart rate, ECG, EEG, Temperature& Arousal, Valence; Cheerful, Happy, Angry, Nervous, Sad; Common BROMP affective categories& -\\
  \hline
  \cite{case} & 30 subjects, Age range: 22-37 years & 8 videos for 4 emotions were used. Duration: 40 minutes & Thought technology sensors& ECG, BVP, EMG, GSR, Resp, Temperature&Amusing, boring, relaxing and scary emotional states& -\\
  \hline
  \cite{clas} & 62 subjects (Mostly Students), Age range: 20-27 years & Data collected using 3 interactive tasks and 2 perspective tasks. Duration: 30 minutes & Shimmer 3& ECG, PPG, EDA and Acc&Valence, arousal and high low concentration, high low cognitive ability& Self-assessment for perspective tasks\\
  \hline
  \cite{amigos} & 40 subjects, Age range: 21-40 years & Two experiments: short video experiment, long video experiment in group and individually. Duration: Short video avg: 1-3 minutes, long video avg: 23.5 minutes & Emotiv EPOC, Shimmer 2R & ECG, GSR and EEG & Valence, arousal & Self-assessment scales, PANAS, personality traits forms, mood assessment\\
  \hline
\end{tabular*}
\end{table*}

Several studies have focused on developing machine learning models for recognizing emotions. Emotion recognition is considered as a promising way to assess wellness. In this section, review of datasets that have been utilized in previous studies to model emotions is provided. The study in \cite{wb_ra_l4} mentions a few video-based datasets that have been used in emotion recognition studies in lab settings. In this study we provide the following information about the datasets: the participating subject group, the data collection scenario, the devices used, the sensors applied, the target variables and the subjective scales used in the study. The following datasets are presented: WESAD \cite{wesad}, DREAMER \cite{dreamer}, K-EmoCon \cite{k_emocon}, CASE \cite{case}, CLAS \cite{clas} and AMIGOS \cite{amigos}. The summarized information from the dataset reviews could be used to formulate the data collection protocols. The review provided the following findings:
\begin{enumerate}
    \item Average number of participants from which the data is collected is 30 and age range being 20-40 years.
    \item In most of the scenarios, the protocol design for data collection consisted of video clips that could possibly trigger emotions such as happy, sad, fear, anger, disgust, joy etc. Other scenarios include stress tests or cognitive tests.
    \item The most common device/combination used to collect the data includes Shimmer devices, Emotive EPOC and Empatica E4. 
    \item For tasks such as emotion recognition both objective and subjective data are crucial. Objective data are collected in the form of physiological responses when the subject is watching a video clip or performing some tasks. Common physiological measures collected include ECG, EEG, EDA or GSR, PPG and mobility (accelerometer data). In addition to objective measurements, subjective data in the form of questionnaire-based scales are used. The most common scales include PANAS and SAM scale.
    \item There are various ways in which the objective data obtained from video tasks can be annotated. The most popular one being the Russell's circumplex model of affect, which represents a two-dimensional emotional space with valence and arousal as axes \cite{wb_ra_l4,circumplex}. Along with the valence-arousal scale, different emotions including amusing, boring, happy, sad, scary etc. are also used.
\end{enumerate}

\subsection{Data analysis}
Data analysis section focuses on the different methods used in the literature to assess the well-being based on either subjective or objective data sources. From the previous subsections, it can be seen that subjective and objective data have been analyzed in a different way in the literature; subjective data have been analyzed mainly using quantitative methods such as statistical analysis, regression analysis or graphical analysis, whereas objective data have been analyzed using machine learning methods. These different types of analysis methods can be described as:
\begin{enumerate}
    \item \textbf{Quantitative analysis}: Quantitative analysis is the process of analyzing the data using statistical methods and tools to identify the trend or pattern in the data. For example, \cite{wb_ra1,wb_ra5}  have applied quantitative analysis methods including regression analysis and statistical analysis using tools such as Statistical Package for Social Sciences (SPSS). They have derived metrics based on frequency or amplitude distribution of the data such as skewness, kurtosis, mean, standard deviation etc. In addition, multiple linear regression analysis was also carried out to identify relationships in the data. Furthermore, researchers in \cite{wb_ra6} have used line plots and scatterplots to analyze speech frequency and duration of conversation with robots. However, in all these studies the data size was limited and data were based on questionnaires.
    
    \item \textbf{Machine learning analysis}: Machine learning analysis utilizes algorithms to model the data and make predictions based on the model. Various machine learning models have been developed to model physiological responses and predict mental states such as emotions, stress or anxiety as wellness indicators. Models such as Hidden Markov Models (HMM), regression analysis, neural networks, Support Vector Machines (SVM), K-nearest neighbor etc. have been used in the literature. In \cite{wb_ra_l4, wb_ra_l1,wb_ra_l3} tree-based methods such as decision trees or random forests as well as boosting models such as XGBoost, light gradient boost or CatBoost have been used to predict emotional states, social anxiety or stress, respectively. These studies have also used other regression-based methods and the SVM algorithm. In \cite{wb_ra_l5}, researchers have developed a system using the k-means algorithm and fuzzy logic to identify human emotions at entrance gates. This model has used real-time body parameters (physiological signals). Other studies have used Convolution Neural Networks (CNN) to model speech signal data for the assessment of different types of mental health disorders such as anxiety, depression, bipolar disorders, obsessive compulsive disorders etc. \cite{mental_health1}. Furthermore, Kalman filter based methods and customizable machine learning with feature selection approach have been used for mental health assessment. 

\end{enumerate}

\subsection{Recommendation systems}
Recommendation systems are software systems that suggest items/content to users based on their interests and behavior. In the context of well-being, a recommendation system can be considered as an application or end result of well-being modeling. Recommendation systems can be developed to suggests recreational activities to the users based on their behavioral analysis and state of well-being. This will help the users to promote their mental, physical and social well-being. Recently, many studies have investigated and developed such recommendation systems that mainly focus on increasing social integration for older people. For example, in \cite{re_sy1} a platform that contributes to physical and social care of older people has been developed. The system developed in \cite{re_sy1} consists of a sensing layer, data processing layer, knowledge gathering layer and smart assistant layer. Similarly, researchers in \cite{re_sy2,re_sy4} have proposed a recommendation system for suggesting recreational activities for elderly people using the information related to demographics, loneliness levels, interests in different activities, disabilities, group size etc. Such system helps assessing, evaluating and improving social engagement of the users. In \cite{re_sy3} a context-aware recommendation system is proposed to improve the mental health of individuals. This system relies on the input from multiple sources such as sensors, user's current state, user profile, mental health intervention database etc., to enhance the recommendation features. Furthermore, the recommendation system proposed in \cite{re_sy3} implements an AI chatbot system that learns from user's feedback and provides recommendations based on the current context. Thus, there exists the possibility to develop AI-driven end-to-end well-being analysis systems that can assess/identify the state of well-being and recommend recreational activity based on user's situation. While mainly targeted for elderly people, such systems can well be developed for individuals of different age groups and different needs.

\section{Future research direction}
Although previous studies have contributed to a great extent in the field of well-being analysis using either real-life environments or lab-based environments, there exists potential to study and refine the data collection, data modeling, data visualization and recommendation setup even further. This section highlights some gaps in the existing literature and proposes a framework to overcome these gaps.
\subsection{Research gaps} \label{subsec:label}
From the comprehensive analysis of literature considering the assessment of well-being through data collection and data analysis, substantial needs for further research can be recognized:
\begin{enumerate}
    \item In previous studies, most of the measurement setups involving physiological responses consider lab environments. However, if well-being analysis is to be applied to everyday activities, natural environments or real-life environments are more relevant. This shows the possibility to develop the recording protocol so that it measures the physiological responses in the real-life setup when the person is involved in some recreational activities.
    \item Well being is a broad concept and involves multiple perspectives such as physical, psychological, social, cognitive, and mental. This requires a range of sensors to be included in the data collection including EDA, EEG, PPG, ECG, Respiration etc. More research is needed to better relate these different variables to the various aspects of well-being.
    \item Interactiveness and transparency in well-being analysis improves the trust and adaptability of the system. A promising opportunity exists in developing visualization tools that could reveal the changes in physiological variables in response to the stimuli posed by real-life recreational activities.
    \item Although previous studies have worked on developing machine learning models for predicting mental states such as anxiety, stress, emotions, mood etc., there is room for improving the performance of these models using more sophisticated methods involving deep learning or transformer models \cite{Transformer_gap}.
\end{enumerate}

\subsection{Execution framework}
\begin{figure*}[h]
\centering
  \includegraphics[width=0.95\textwidth]{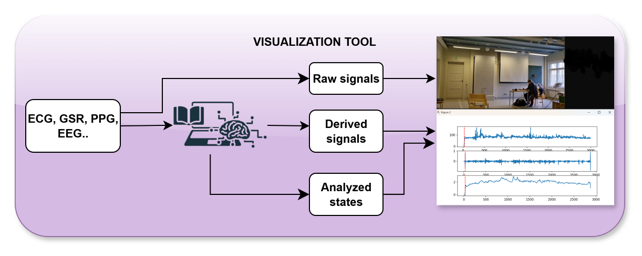}
  \caption{Execution framework}
  \label{fig:execution_framework}
\end{figure*}

In this section, we propose an execution framework which redefines the data collection, data visualization and data analysis to obtain better results for enhancing the well-being using AI-driven approaches. Figure \ref{fig:execution_framework} presents the general concept of the framework. The flow of events during a recreational activity is presented in the upper part of the visualization window. This can be in the form of a video recording (in the case of a theater performance or a concert, for example), a recording of a computer screen (in the case of computer gaming, for example), an audio recording, or any other relevant information. Below that, the time series data are displayed. This can contain:
\begin{enumerate}
    \item Raw time series from physiological sensors such as ECG, EEG, GSR, PPG or Accelerometer 
    \item Derived variables such as Heart Rate derived from ECG or PPG, or
    \item Model outputs representing a prediction of some mental state such as stress, anxiety, emotions or mood, for example.
\end{enumerate}

\begin{figure}
  \includegraphics[width=\columnwidth]{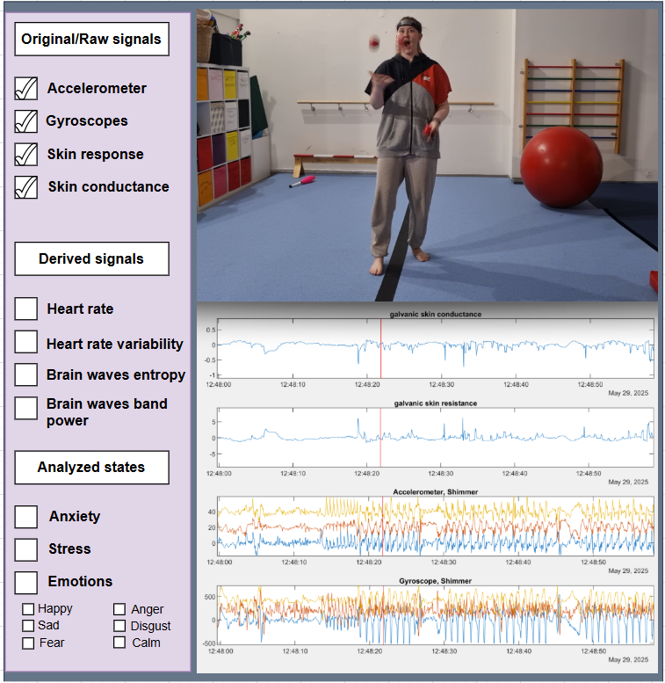}
  \caption{Simulation visualization tool}
  \label{fig:simualtionadvanced}
\end{figure}

The strategy is to visualize the signals along with the video or a recording of the environment in which the data were acquired. In Figure \ref{fig:simualtionadvanced} the sensor data were taken from the performer presenting a joggling session. The user interface presents the choice of the signals to be viewed and the data window shows the video recording and the corresponding signals, in this case the signals from the accelerometer and gyroscope together with skin resistance and skin conductance from the GSR sensor. The red line acts as an indicator and is synchronized with the video. The simulation shows only a few signals, most of which are raw signals. However, the advanced version of the visualization tool includes the options to visualize derived signals and analyzed states along with raw signals (as shown in Figure \ref{fig:simualtionadvanced}).  The derived signals that could be visualized include heart rate, heart rate variability, brain wave entropies or brain wave band power. The analyzed states that could be visualized include anxiety, stress and various emotions. 

The users can benefit from the proposed well-being analysis tool in the following manner:
\begin{enumerate}
    \item It helps in visualizing the deviations/patterns form the normal range, in this case emotions.
    \item It helps in recognizing moments of stress, anxiety, or arousal levels, enabling self-regulatory behavior. 
    \item Real-time outputs through such system is essential in biofeedback and mental health applications.
    \item The tool is of great interest to the performers and artists to find out the reaction of the audience to the various scenes and events.
\end{enumerate}

\section{Discussion}
This study examined the state-of-the-art methods for developing AI-driven models to enhance the well-being using physiological responses. In this review, we define three principal concepts, well-being, recreational activities and physiological responses, as well as the relationship between them. We investigated previous studies for data collection environments, data collection setup, sensors, devices, analysis methods, and evaluation criteria for detecting/predicting well-being indicators. This review also examined the various real-life and lab environments that have been used in data collection scenarios along with the type of data collected, for well-being analysis. This comprehensive review also explored the datasets, devices and sensors used in the previous studies for well-being analysis. The study identified various well-being indicators such as anxiety, depression, stress, emotions, mood changes etc. However, the extensive review conducted here assisted in identifying the existing areas that require more research for enhanced AI driven well-being analysis. The significant potential exists in the field of improving the data collection setup for real-life environments and data analysis, predominantly machine learning analysis. In addition to the comprehensive review, the study proposed an execution framework to work on the unexplored areas. The execution framework proposes a data collection protocol that includes multiple physiological responses such as heart rate, skin conductance, heart rate variability, activity data in lab and real-life environments. Another part of the execution framework aims to develop machine learning models that can recognize behavioral patterns specially in real-life environments. The proposed setup also suggests developing a real-time visualization framework for identifying deviation in physiological responses. This will not only enhance the transparency but also helps in improving mental health application. 
From a clinical perspective, these findings extends to preventive care. By identifying factors that  influence well-being, clinicians in primary care and community settings may be better equipped to implement early, low-intensity interventions to reduce the progression to more severe mental health conditions.

\section{Conclusion}
In conclusion, this review demonstrates the factors influencing well-being and highlights its relevance for both research and clinical practice. The findings illustrated the importance of well-being assessment and how addressing well-being can help in enhancing user engagement and improving quality of life. Additionally, the proposed framework suggested that continued research is needed to strengthen the evidence base and develop effective, scalable AI-driven approaches to enhancing well-being.

\bibliographystyle{IEEEtran}
\bibliography{Bibliography}{}

\end{document}